\documentclass[twocolumn,showpacs,amsmath,amssymb,prl,a4paper]{revtex4}
\usepackage{graphicx}
\usepackage{dcolumn}
\usepackage{bm}
\newcommand{\PP}{\mathcal{P}}
\newcommand{\TT}{\mathcal{T}}

\begin{document}

\title{Formation and annihilation of laser light pulse quanta in thermodynamic-like pathway }

\author{Boris Vodonos, Rafi Weill, Ariel Gordon, Alexander Bekker, Vladimir Smulakovsky,
Omri Gat and Baruch Fischer}
\affiliation{Department of Electrical Engineering, Technion, Haifa
32000, Israel }

\date{\today}

\begin{abstract}
We present a theoretical and experimental study of multiple pulse
formation in passively mode-locked (PML) lasers. Following a
statistical mechanics approach, the study yields a
thermodynamic-like ``phase diagram" with boundaries representing
cascaded first order phase transitions. They correspond to abrupt
creation or annihilation of pulses and a quantized RF power
behavior, as system parameters (noise and/or pumping levels) are
varied, in excellent accordance with the experiments. Remarkably,
individual pulses carry an almost constant quantum of energy.
\end{abstract}


\maketitle

Lasers operating in a pulse regime can develop more than one pulse
running in their cavities \cite{Richardson}. That is, at any given
instant, the optical energy is sharply concentrated around several
points along the cavity rather than at one point (a single pulse
regime) or evenly distributed over the cavity (continuous wave
operation). This multipulse regime can be triggered, for example,
by a suitable form of absorptive nonlinearity. These lasers
exhibit rich highly-nonlinear and complicated dynamics, the
modeling of which
\cite{Akhmediev,Komarov,Lederer,Namiki,Kalashnikov} requires
inclusion of nonlinearities of higher order than the common
quartic (Landau-Ginzburg-like) term.

A remarkable feature often exhibited by these lasers is the
quantization of the pulse energies \cite{Grudinin,Matsas}. That
is, all pulses possess nearly fixed energy, which is almost
independent of, for example, the pumping power. These pulse quanta
can travel with respect to one another, attract or repel each
other and form ordered structures, such as couples or bunches.
These phenomena have been lately receiving increasing attention
\cite{Kutz,Menyuk,Grudinin_Gray,Tang,Grelu,Seong,Gutty,Soto-Crespo}.

Recently it has been shown \cite{GFPRL} that the interplay between
nonlinearities in the laser on one hand and noise, such as the
inevitable spontaneous emission noise, on the other hand, can
account for formation and destruction of a single pulse in PML
lasers. This discontinuous formation and destruction of pulses
was identified as a first order phase transition, governed by the
balance between interaction (due to the nonlinearity) and entropy
(associated with noise). The study is taken beyond the simple
\textit{mechanics} view of pulses and modes in lasers, to a
\textit{statistical-mechanics} picture where a central role is
played by basic thermodynamic quantities, such as "temperature"
(noise), entropy and free energy.

In this Letter we present new theoretical and experimental results
concerning the thermodynamics of pulses in lasers. We focus on
the very creation and annihilation of pulses in a multi-pulse
regime as the pumping and the intracavity noise level are varied.
We find that as noise is increased and/or pumping is decreased,
pulses successively disappear, one by one, in what is shown to be
a cascade of first order phase transitions. This novel
thermodynamic multi-phase light system gives rise to a quantized
RF power effect. The theory relies on an equilibrium-like
statistical mechanics model that is solved analytically. The
qualitative and quantitative agreement between theory and
experiment is excellent.

The dynamics of the (classical) electric field in a PML laser is
commonly described by the complex Ginzburg-Landau equation, which
can also include high order nonlinearities \cite{Akhmediev},
especially in the context of multiple pulse operation. In our
study we wish to include the effect of noise as well, which would
render the above description difficult to handle analytically or
numerically. We therefore construct here a much simpler model of
the dynamics, which is capable, as the experiment demonstrates, of
capturing the key features of noise-dependent formation and
destruction of pulses.

We divide the cavity to $N$ equal intervals, such that $N/\tau$
($\tau$ is the cavity roundtrip time) is of the order of the
bandwidth of the laser, which means that the duration of a pulse
is of order $\tau/N$. Then we write an equation of motion for the
energy $x_m$ at the $m$-th interval:
\begin{equation}\label{EOM:pulse}
\frac{dx_m}{dt}=s(x_m)x_m + gx_m+ \sqrt {x_m}\Gamma_m(t)
\end{equation}
$s(x)$ is the effective nonlinear gain for the energy in the
interval, $t$ is the long scale time variable over which the laser
evolves between roundtrips, $g$ is the overall net gain
(originating from the slow amplifier and effective linear
losses), $\Gamma_m$ is real white Gaussian noise satisfying
$\langle\Gamma_m(t)\Gamma_m(t')\rangle=T\delta_{mn}\delta(t-t')$
and the last term is in the Stratonovich interpretation
\cite{Risken}. Properly modifying $g$ one can always set $s(0)=0$.

When the $m$-th interval is occupied by a pulse, an equation of
the form of Eq.~(\ref{EOM:pulse}) has been established in
previous studies and both $s(x)$ and the form of the noise term
given in Eq.~(\ref{EOM:pulse}) were obtained: The latter for
solitons \cite{Mecozzi} and the former also for other types of
pulses \cite{Namiki}. If on the other hand the $m$-th interval
belongs to the continuum, its energy is small enough such that
nonlinearities are negligible. Then neglecting $s(x_m)$ and the
dependence of $g$ on $x_m$ in Eq.~(\ref{EOM:pulse}), we are left
with the same equation of motion that we would obtain for the
energy of a sample of a band-limited function under the effect of
white Gaussian noise.

We assume that the interaction between different intervals is
weak enough such that the equations of motion of $x_m$-s are not
coupled, apart from a global constraint of constant total energy
$\PP$, introduced by choosing $g$ as the appropriate Lagrange
multiplier \cite{GFPRL,GGF03}.

While in previous energy rate equation studies \cite{Namiki} the
energy of the continuum was represented by a single degree of
freedom, in our study this is not appropriate. When noise is
present \emph{the continuum carries essentially all the entropy}
\cite{GGF03} and hence it is crucial that it is represented by
many degrees of freedom.

Following the same steps as in Refs. \cite{GFPRL, GFOC, GGF03},
one can show that the invariant measure imposed by
Eq.~(\ref{EOM:pulse}) is a Gibbs distribution described by the
partition function
\begin{equation}\label{eq:Z-def}
Z_N(T,\PP)=\int  dx e^{\frac {1}{ T} \sum\limits_{m=1}^N
S(x_m)}\delta \left(\sum_{m=1}^N x_m-\PP\right)
\end{equation}
where $S(x)=\int_0^x s(x')dx'$ and $dx$ denotes integration with
respect to all $x$-s from zero to infinity.

>From Eq.~(\ref{eq:Z-def}) it is clear that the $x_m$-s are
bounded by $\PP$. If $s(x)$ is an increasing function of $x$ for
$x<\PP$, which can happen when a saturable absorber is present
and $\PP$ is small enough, pulses usually do not split. Pulse
splitting occurs typically when $s(x)$ is an increasing function
at $0<x<x_s$ for some $x_s$ and is a \emph{decreasing} function
for $x>x_s$ \cite{Haus,Kaertner,Doerr}, which is the situation in
for example in additive pulse mode locking \cite{Haus}: It is
intuitively clear that one pulse with a energy much higher than
$x_s$ will be less favorable than many pulses with a energies of
order $x_s$. We henceforth assume the above described structure of
$s(x)$, and show that when $\PP\gg x_s$, Eq.~(\ref{eq:Z-def})
predicts formation of a variable number of pulse ``quanta",
i.~e.~pulses with nearly constant energy \cite{Doerr}. Pulse
quanta are spontaneously and abruptly created and annihilated
when $\PP$ and $T$ are varied.

The model studied in Ref.~\cite{GGF03} is a special case of (2),
in which the function $S$ is quadratic. There, however, the model
was derived under somewhat different conditions than in the
present work. Thus, if $s$ is chosen linear at the origin ($S$ is
quadratic), the present results recapture those of
Ref.~\cite{GGF03} for small $\PP$, but we expect that they do not
describe the actual behavior of the multipulse laser at small
powers.

\begin{figure}[htb]
\hbox{\includegraphics[width=4.25cm]{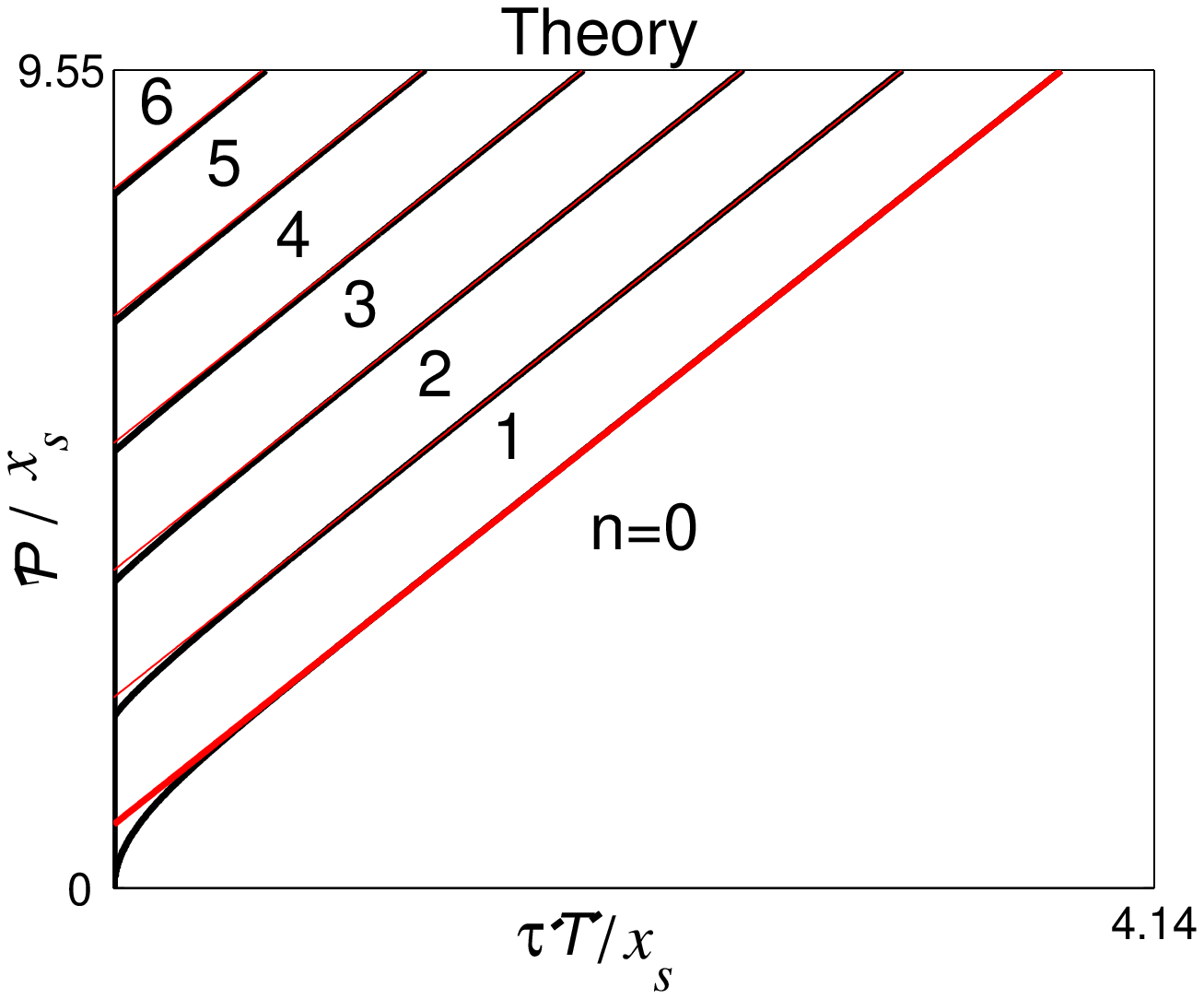}
\hskip0.1cm\includegraphics[width=4.25cm]{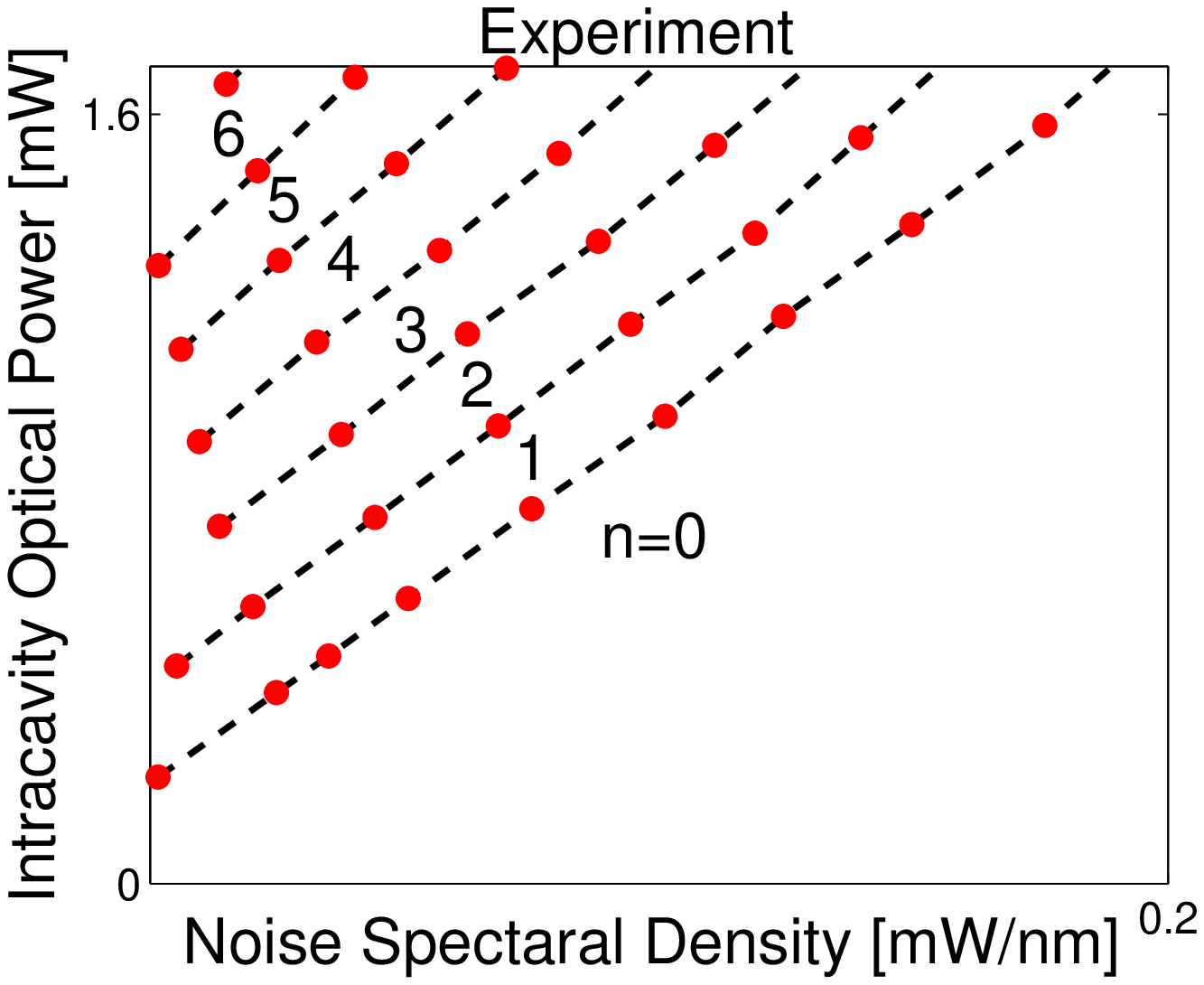}}
\caption{Experimental (right) and theoretical (left) phase
diagrams. The theoretical graph shows the number of pulses as a
function of the intracavity energy $\PP$ and the total noise
power $\TT$, Eq.~(\ref{eq:transmissivity}). The curves of
discontinuity have the thermodynamic-like meaning of first order
phase transitions. The straight lines are asymptotes of the
transition lines for $\PP\gg x_s$ (Eq.~(\ref{eq:transcondd})).
\label{fig:phase}}
\end{figure}

In the thermodynamic theory of our system, the formation of a
single pulse is a first order phase transition
\cite{GFPRL,GFOC,GGF03}, and the formation of multiple pulse
configurations is a first order transition between different
ordered ``phases'', reminiscent of structural phase transitions in
solids \cite{structural}. To show the remarkable resemblance to
the thermodynamic phase picture with the predictive power of the
statistical-mechanics theory, we refer right at the beginning to
Fig.~\ref{fig:phase}. The latter shows the theoretical ``phase
diagram" derived below from Eq.~(\ref{eq:Z-def}), and also
experimental measurements as described below. Each phase is
labelled by the number of pulses.

We proceed to outline the analysis of the statistical mechanics
model of Eq.~(\ref{eq:Z-def}), which is similar one given in
detail in Ref.~\cite{GGF03}. We make the physically appropriate
assumption that $N\gg1$, i.~e.~that $\tau$ is much longer than
the duration of a single pulse (or equivalently the number of
modes in the laser, which is of order $N$, is large). This
amounts to taking the thermodynamic limit in Eq.~(\ref{eq:Z-def}).

We start by assuming a specific asymptotic form (as $N\to\infty$)
for the partition function $Z_N^{(m)}$ of configurations with $m$
pulses or less
\begin{equation}\label{eq:AMF}
Z_N^{(m)}(T,\PP) = A_N^{(m)}(T,\PP)e^{-N(F_m(T,\PP)-1)}
\end{equation}
where $A_N^{(m)}(T,\PP)$ is sub-exponential in $N$ and
$F_m(T,\PP)$, the free energy, is the global minimum of
\begin{equation}\label{eq:f}
f_m(x_1,..,x_m)=-\frac{1}{NT}\sum_{j=1}^m
S(x_j)-\ln\Big(\PP-\sum_{j=1}^m x_j\Big)
\end{equation}
for nonnegative $x$ values such that $\sum x_j\leq \PP$. Let $n$
be the smallest $m$ for which the minimum of $F_m(T,\PP)$ with
respect to $m$ is attained, and let $X_1,...X_n$ be the
corresponding minimizer. Our statement then is that $Z_N$
approaches $Z_N^{(n)}$ in the thermodynamic limit, and in
particular $n$ is the number of the pulses per roundtrip and
$X_1,...X_n$ are their energies.  If $n$ and the $X_j$-s are to
have a finite thermodynamic limit, $T$ has to scale like $1/N$,
so that $\TT\equiv NT$, the total power of noise, has a finite
limit. The necessity of rendering the parameters of the system
$N$ dependent is discussed in \cite{GGF03}.

Eqs.~(\ref{eq:AMF},\ref{eq:f}) follow from Eq.~(\ref{eq:Z-def}) if
one assumes that as $N\to\infty$ all but a finite number $m$ of
the $x$ variables are $O(1/N)$.  This is justified by a rigorously
controlled approximation in $1/N$ based on a recursive equation
for $Z_N$, as in Ref.~\cite{GGF03}. The upshot is that the task of
calculating thermodynamic properties is reduced to that of
finding the minima of $f$ for different values of $\PP$ and $\TT$.
In this manner, not only the number of pulses per roundtrip is
obtained, but many other quantities of interest, such as the order
parameter
\begin{equation}\label{eq:Q}
Q=\sum_{j=1}^N\left<x_j^2\right> =\sum_{j=1}^nX_j^2
\end{equation}
which is proportional to the experimentally measurable RF power
of the photocurrent \cite{experiment}.

Since $m$ is finite, finding the minimum of $f_m$ for a specific
choice of $s(x)$ is straightforward, although in general it
includes a numerical solution of a set of transcendental
equations. However for the structure of $s(x)$ we consider in
this Letter, asymptotic expressions for $\PP\gg x_s$, which is
the domain of the multipulse regime, can be obtained. Standard
minimization techniques show then that the minima of $f_m$ are
usually obtained at two types of configurations: the first has
$n\leq m$ nonzero $x$ values all having the same value $X$,
i.~e.~configurations of $n$ equipotent pulses, and the second,
which is much rarer, is $n-1$ values of $x=X$ and one additional
value $x=X'<X$. A sufficient condition for excluding the second
type of solutions is that $s(x)$ increases at least as fast as it
decreases:
\begin{equation}\label{c}
s(x_1)=s(x_2),\,x_1<x_2 \Rightarrow |s'(x_1)|\geq|s'(x_2)|,
\end{equation}

Assuming the first type of solutions, the problem reduces to the
minimization of the function
\begin{equation}f(n,x)= -\frac{nS(x)}{\TT}-\ln(\PP-nx)\ ,\label{eq:fnx}\end{equation}
with respect to two variables, $n$ (integer) and $x$
($0\leq~x~\leq~\PP/n$). The minimizer $X$ satisfies
\begin{equation} s(X)=\frac{\TT}{\PP-nX}\ .\label{eq:X}\end{equation}
Clearly $X$ is the common pulse energy and the order parameter is
$Q=nX^2$.

The asymptotic regime relevant to a multi-pulse operation,
$\PP\gg x_s$, is most readily analyzed by considering the
minimization of the same function $f$ appearing in Eq.
(\ref{eq:fnx}) but with $n$ replaced by a real valued variable
$\nu$. Minimizing with respect to $x$ and $\nu$ together
immediately gives $X=X_*$, where $X_*$ is the solution of
\begin{equation}\label{eq:magic}
S(X_*)=X_*s(X_*)\ .
\end{equation}
$X_*$ is the quantized pulse energy. From the assumptions made
above on $s(x)$ it follows that $X_*$ is unique as long as it
exists, and $X_*>x_s$ (remarkably enough). We use the notation
$s_*=s(X_*)$, $S_*=S(X_*)$. Then writing $X=X_*+\delta$ and
$\nu=n+\{\nu\}$ with $n$ integer and $|\{\nu\}|\leq\frac12$, and
putting back in Eq. (\ref{eq:X}) gives
\begin{equation}
\frac{\delta}{X_*}=\frac{\{\nu\}s_*}{s'(X_*)(\PP-\nu X_*)-ns_*}
\end{equation}
if $\delta\ll X_*$. Since $s'(X_*)<0$ while $s_*>0$, indeed
$\delta\ll X_*$ provided either $\PP-nX\gg X_*$ or $n\gg1$. At
least one of them is certain to hold whenever $\PP\gg X_*$ This
asymptotic region corresponds to the upper part of
Fig.~\ref{fig:phase}, and is characterized by a nearly quantized
pulse energy.

The phase diagram in the quantized pulse regime is very simple:
The transition from an $(n-1)$-pulse configuration to an
$n$-pulse configuration occurs approximately when $\nu$ is half
an odd integer. Using Eq. (\ref{eq:X}) once more gives the
transition temperature $\TT_n(\PP)$
\begin{equation}\label{eq:transcondd}
\TT_n(\PP)=s_*\PP-(n-1/2)S_*\ ,
\end{equation}
i.~e.~the transition curves are approximately equally spaced
straight lines, as seen in Fig~\ref{fig:phase}.

\begin{figure}[htb]
\hbox{\includegraphics[width=4cm]{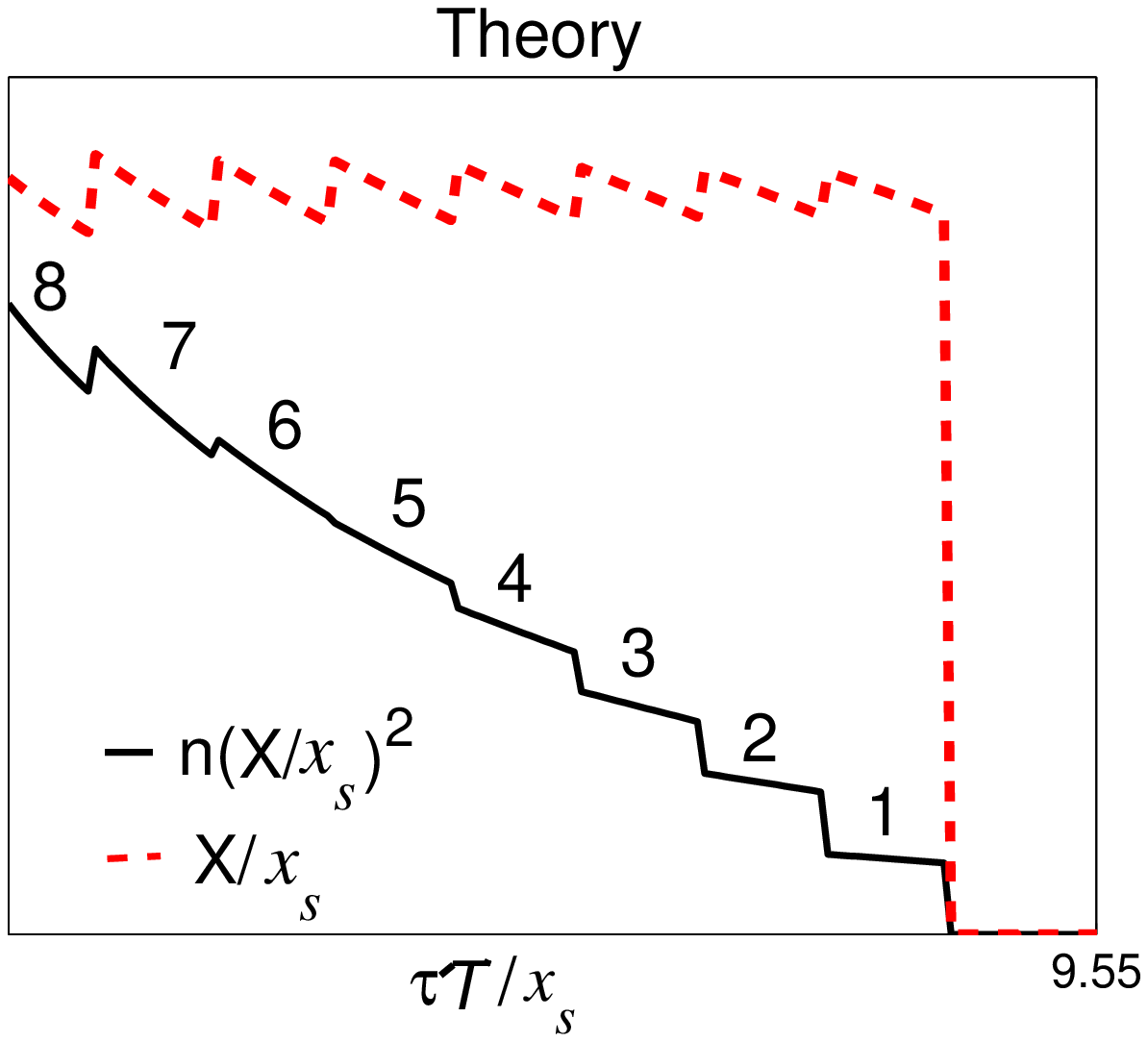}
\hskip0.1cm\includegraphics[width=4.45cm]{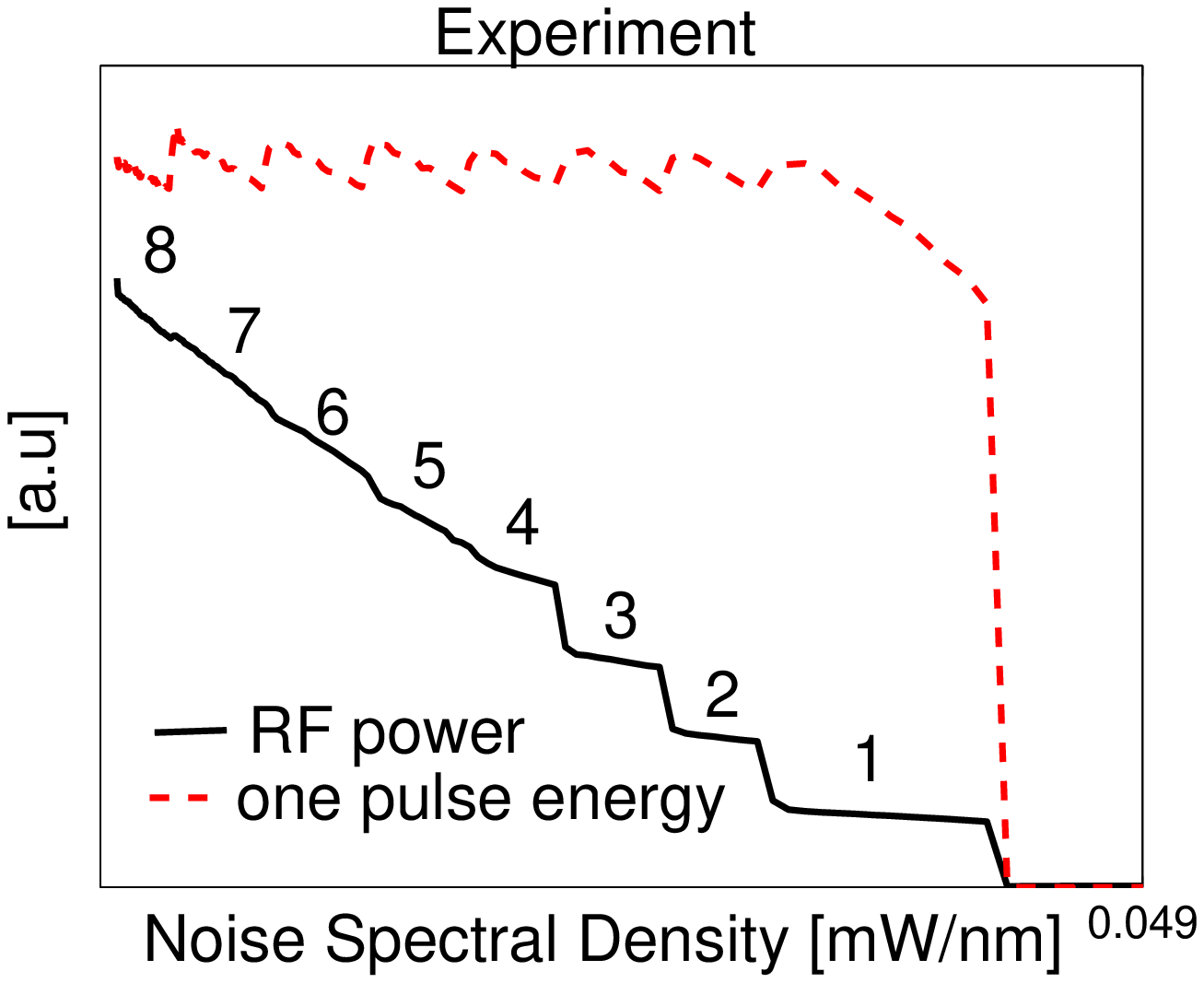}}

\caption{Experimental (right) and theoretical (left, with
Eq.~(\ref{eq:transmissivity})) plots of the order parameter $Q$
(solid line) and the mean pulse energy (dashed line) as functions
of the noise spectral power, for a fixed intracavity energy.
\label{fig:quanta}}
\end{figure}

\begin{figure}[htb]
\hbox{\includegraphics[width=4.05cm]{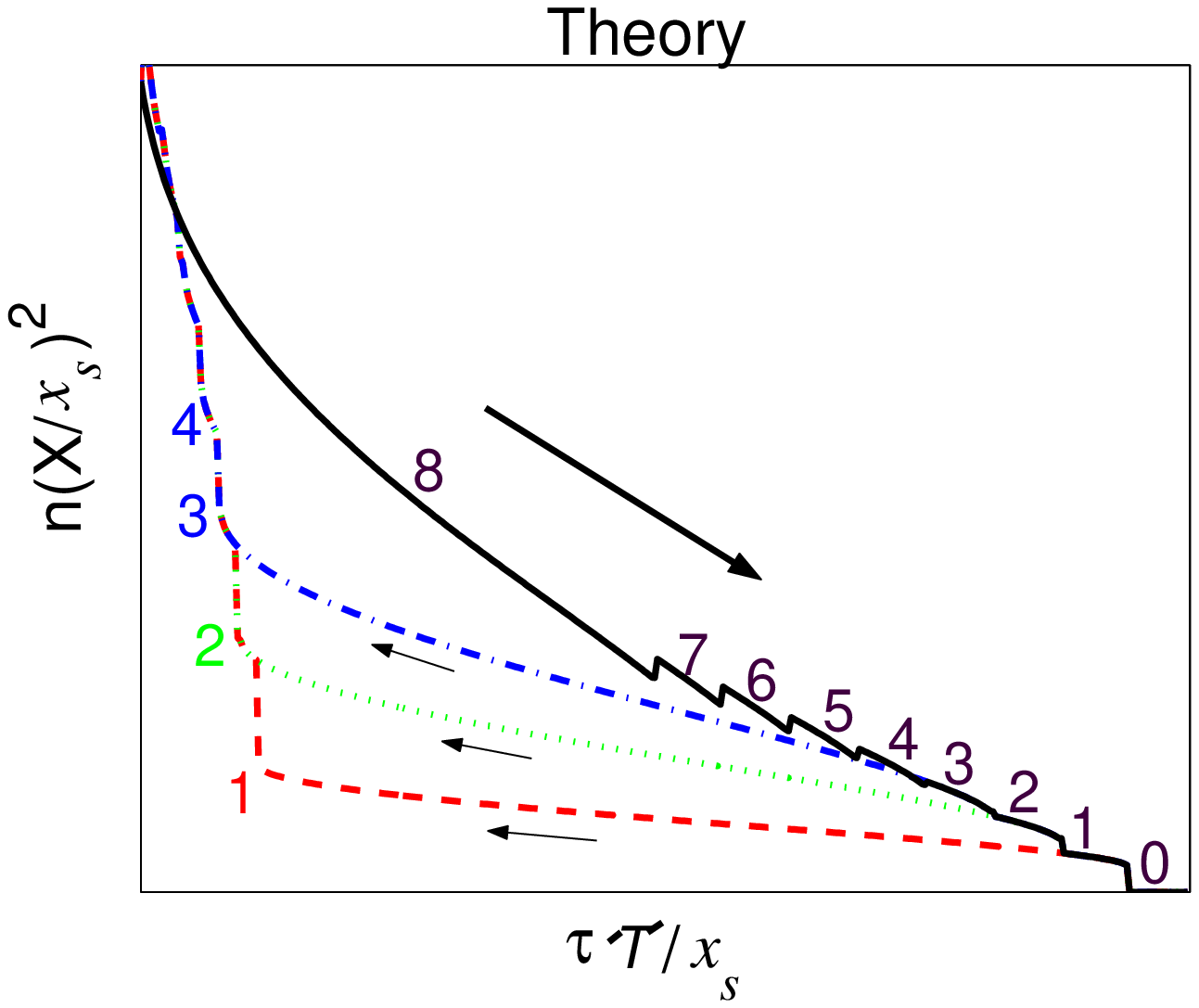}
\hskip0.1cm\includegraphics[width=4.25cm]{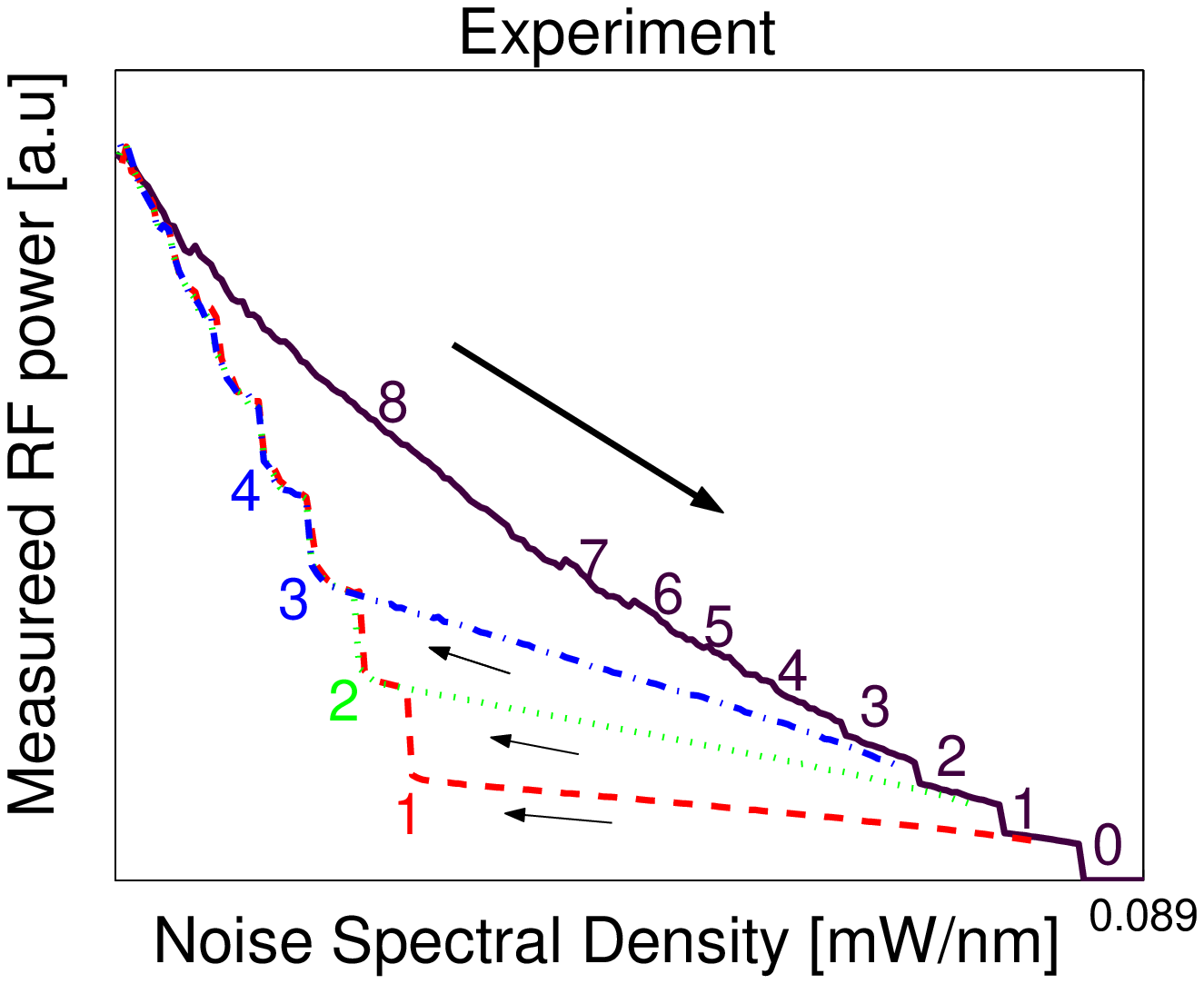}}
\caption{Experimental (right) plot of the order parameter $Q$ as
a function of increasing (solid line) and decreasing (broken
lines) noise spectral power, demonstrating three different
hysteresis paths, with excellent agreement to the theoretical
plot (left), with the choice $s(x)=a_1x-a_3x^3+a_5x^5$.
A motion picture demonstration of the experimental results can be views in EPAPS document no [XXXX].
\label{fig:order}}
\end{figure}

In order to illustrate our results we chose
\begin{equation}\label{eq:transmissivity}
s(x)=\tau^{-1}\sin\left(\frac{\pi x}{2x_s}\right)
\end{equation}
The motivation of this choice is the sinusoidal transmissivity in
additive pulse mode locking \cite{Haus}, and although $s(x)$, the
net gain experienced by a pulse, somewhat differs from the bare
transmissivity \cite{Namiki}, we disregard this difference here,
since the multiple pulse regime is anyway governed by the
neighborhood of $X_*$. Results for Eq.~(\ref{eq:transmissivity})
are shown in Fig.~\ref{fig:phase} for the number of pulses and in
Figs.~\ref{fig:quanta} and \ref{fig:order} for the order
parameter $Q$.

Our experimental study was conducted on a setup similar to a
recently reported one \cite{experiment}, where the first order
phase transition associated with the formation and destruction of
a single pulse was lucidly demonstrated. It consists of a fiber
ring laser with PML by nonlinear polarization rotation technique
\cite{Haus,Matsas}, with amplified spontaneous emission noise
injected from an external source, in order to have direct control
over the spectral power of the additive noise in the cavity. Here
we used a shorter laser with a roundtrip time of 100 nanoseconds,
corresponding to approximately 20 meters of total cavity length,
including 4.3-m long erbium-doped fiber amplifier with small
signal gain of 6dB/m. By proper adjustment of the polarization
controllers (PCs) PML operation was established with generation
of sub-picosecond pulses. As observed in a variety of PML fiber
lasers \cite{Richardson,Grudinin,Matsas,Haus}, excessive pumping
(above the self-starting power threshold) led to the formation of
multiple pulses per roundtrip that in general were randomly
distributed over the cavity. However, for certain positions of
PCs, stable bunches of nearly identical pulses were formed, with
approximately constant spacing between adjacent pulses (that
ranged from a few to hundreds of picoseconds, depending on PCs
positions).

As the noise or pumping levels were varied, two types of responses
of the pulse bunch were observed: variations in the spacing
between adjacent pulses and variations in the pulse energy.
Therefore, depending on the position of the PCs, three distinct
regimes of bunched pulse operation were obtained. The first and
the most common was the regime where both types of response were
observed. In the second regime the multi-pulse bunch contracted
or expanded while pulse energies remained constant, and the third
regime was characterized by a fixed bunch pattern while pulse
energies were varied.

Fig.~\ref{fig:phase} shows the experimental phase diagram measured
as follows:  for several pumping powers the injected noise level
was raised gradually from zero, the pulses disappeared one by one
and the transition ``temperatures" and average output optical
powers were recorded. Such a behavior was previously observed as
the pumping power was decreased \cite{Matsas}. The experimental
results presented in Fig.~\ref{fig:phase} were obtained at the
first operation regime but the structure of the phase diagram was
found to be identical in all the regimes mentioned above.
Fig.~\ref{fig:phase} demonstrates good agreement between theory
and experiment.

Theoretical and experimental plots of the order parameter $Q$
(Eq.~(\ref{eq:Q})) and the energy per pulse as function of the
injected noise level (gradually increased from zero) for a fixed
pumping power are shown in Fig.~\ref{fig:quanta}. Experimentally
they were obtained by measuring the laser output with a fast
photodiode and an RF power meter \cite{experiment} or a sampling
oscilloscope (all having 50GHz bandwidth) correspondingly. The
pulse energy is nearly constant, with deviations of about 5\%.
These deviations are well described by the theory. The results of
Fig.~\ref{fig:quanta} were
obtained at the third operation regime (constant spacing regime
described above, where pulse energy changes are most pronounced).

Fig.~\ref{fig:order} shows additional theoretical and experimental
plots of the order parameter dependence on noise for both
increasing and decreasing of the noise level. Typically to first
order phase transitions, the system exhibited hysteresis: The
number of pulses at any point ($\TT,\PP$) depends on the precise
path that led to it. In particular, increasing $T$ leads to a
different $Q(T)$ curve than decreasing it, as seen in
Fig.~\ref{fig:order}. The hysteresis can be theoretically
estimated using an analog to the Arrhenius formula: the system
dwells in a meta-stable phase, corresponding to a local minimum of
$f$, as long as this minimum is ``deep". Since the lifetime of a
metastable phase is exponential in the barrier surrounding the
corresponding local minimum of $f$, it can be much longer than the
time over which the system parameters are varied.

We conclude by noting how powerful the combination of
statistical-mechanics and laser physics can be, leading us to a
new view and findings that can be significant to both fields.

\begin{acknowledgments}
We thank Shmuel Fishman for helpful discussions. This work was
supported by the Israel Science Foundation (ISF) founded by the
Israeli Academy of Sciences.
\end{acknowledgments}

\end{document}